\documentclass[sigconf,9pt,nonacm]{acmart}

\usepackage{algorithm}
\usepackage[noend]{algpseudocode}
\usepackage{graphicx}
\usepackage{adjustbox}
\usepackage{caption}
\usepackage{textcomp}
\usepackage{amsthm}
\usepackage{verbatim}
\usepackage{subcaption}
\usepackage{multirow}
\usepackage{booktabs}
\usepackage{makecell}

\usepackage{booktabs}
\usepackage[table]{xcolor}
\usepackage{array}
\usepackage{nicematrix}
\usepackage{tikz}
\usepackage{threeparttable}
\usepackage{tablefootnote}
\usepackage{enumitem}
\usepackage{dblfloatfix}

\copyrightyear{2026}
\acmYear{2026}
\setcopyright{cc}
\setcctype{by}
\acmConference[DAC '26]{63rd ACM/IEEE Design Automation Conference}{July 26--29, 2026}{Long Beach, CA, USA}
\acmBooktitle{63rd ACM/IEEE Design Automation Conference (DAC '26), July 26--29, 2026, Long Beach, CA, USA}
\acmDOI{10.1145/3770743.3804129}
\acmISBN{979-8-4007-2254-7/2026/07}

\def\BibTeX{{\rm B\kern-.05em{\sc i\kern-.025em b}\kern-.08em
    T\kern-.1667em\lower.7ex\hbox{E}\kern-.125emX}}
\begin{document}

\fancyhead{}
\renewcommand{\headrulewidth}{0pt}

\title{IMPart: Integration of Memetic Operations into Multi-Level Framework for Large-$k$-Way Hypergraph Partitioning}

\author{Yugao Zhu,\ Zhicheng Guo, Shang Liu,  Mengming Li, Jing Wang, Zhiyao Xie}
\authornote{corresponding author}
\affiliation{%
  \institution{Hong Kong University of Science and Technology}
  \country{\{yzhuel, zguobx, sliudx, mengming.li, jwangjw\}@connect.ust.hk, eezhiyao@ust.hk}
}

\begin{abstract}    
    \label{sec:abs}
    The problem of $k$-way hypergraph partitioning is fundamental with significant applications in various fields, including VLSI design and scientific computing. State-of-the-art hypergraph partitioners commonly employ a multi-level framework encompassing coarsening, initial partitioning, uncoarsening, and refinement phases. However, many existing methods do not scale well to problems requiring a large number of partitions (i.e., large~$k$). In pursuit of exceptionally high solution quality, existing memetic approaches often execute their two key operations, recombination and mutation, by invoking separate, standalone multi-level partitioners. This design choice, however, renders them significantly more time-consuming than standard multi-level partitioners. To make such memetic approaches more practical, we propose an advanced memetic framework, IMPart, which introduces novel recombination and mutation operators and integrates them directly into the uncoarsening phase of a single multi-level framework. This transforms the local searches of different granularities in the traditional multi-level framework into a sophisticated, collaborative search. Experimental results on multiple standard benchmarks demonstrate our framework more effectively escapes local optima and explores the global solution space for higher-quality solutions, substantially outperforming all existing hypergraph partitioners for large-$k$-way hypergraph partitioning. Our framework highlights a new paradigm for the development of advanced hypergraph partitioners.

\end{abstract}

\maketitle

\newcommand*\circled[1]{\tikz[baseline=(char.base)]{
    \node[shape=circle, fill=black, text=white, inner sep=0.8pt, font=\sffamily\bfseries\small] (char) {#1};}}

\definecolor{ForestGreen}{rgb}{0.13, 0.55, 0.13}

\newenvironment{green}
  {\color{ForestGreen}}
  {}

\section{Introduction}
    \label{sec:intro}
    Hypergraph partitioning is an important problem with widespread applications across various domains, including VLSI design and scientific computing \cite{papa2007hypergraph}. $k$-way hypergraph partitioning is an NP-hard problem \cite{hartmanis1982computers} that aims to divide the nodes of a hypergraph into $k$ disjoint partitions while minimizing the number of hyperedges cut (cut-size). Over the past few decades, a rich array of heuristic algorithms has emerged, including \emph{spectral methods} \cite{zhou2006learning,bustany2022specpart,bustany2023k}, \emph{flow methods} \cite{gottesburen2019evaluation,gottesburen2022parallel}, \emph{local search methods} (e.g., Kernighan-Lin (KL) \cite{kernighan1970efficient} and Fiduccia-Mattheyses (FM) \cite{fiduccia1988linear}), \emph{memetic (genetic) methods} \cite{areibi2004effective,andre2018memetic}, deep learning-based methods~\cite{liang2024medpart,chen2024hypergraph}, and the highly successful \emph{multi-level framework} \cite{karypis1997multilevel,schlag2023high,ccatalyurek2011patoh}.  

The \emph{multi-level framework} is widely regarded as the most successful approach to hypergraph partitioning \cite{sybrandt2020hypergraph,acikalin2022multilevel,gottesburen2021scalable,gottesburen2024scalable}. It forms the basis for leading partitioners like hMETIS \cite{karypis1997multilevel} and KaHyPar \cite{schlag2023high}. This framework operates in three stages: (1) \textbf{Coarsening:} The hypergraph is iteratively coarsened by merging tightly connected nodes. (2) \textbf{Initial Partitioning:} A balance-constrained partition is computed on the coarsest hypergraph. (3) \textbf{Uncoarsening and Refinement:} The partition is projected back through intermediate levels to the original hypergraph, with local search (e.g., FM) applied at each level to refine the solution and reduce the cut size.

\begin{figure}[!t]
    \centering
    \includegraphics[width=\linewidth]{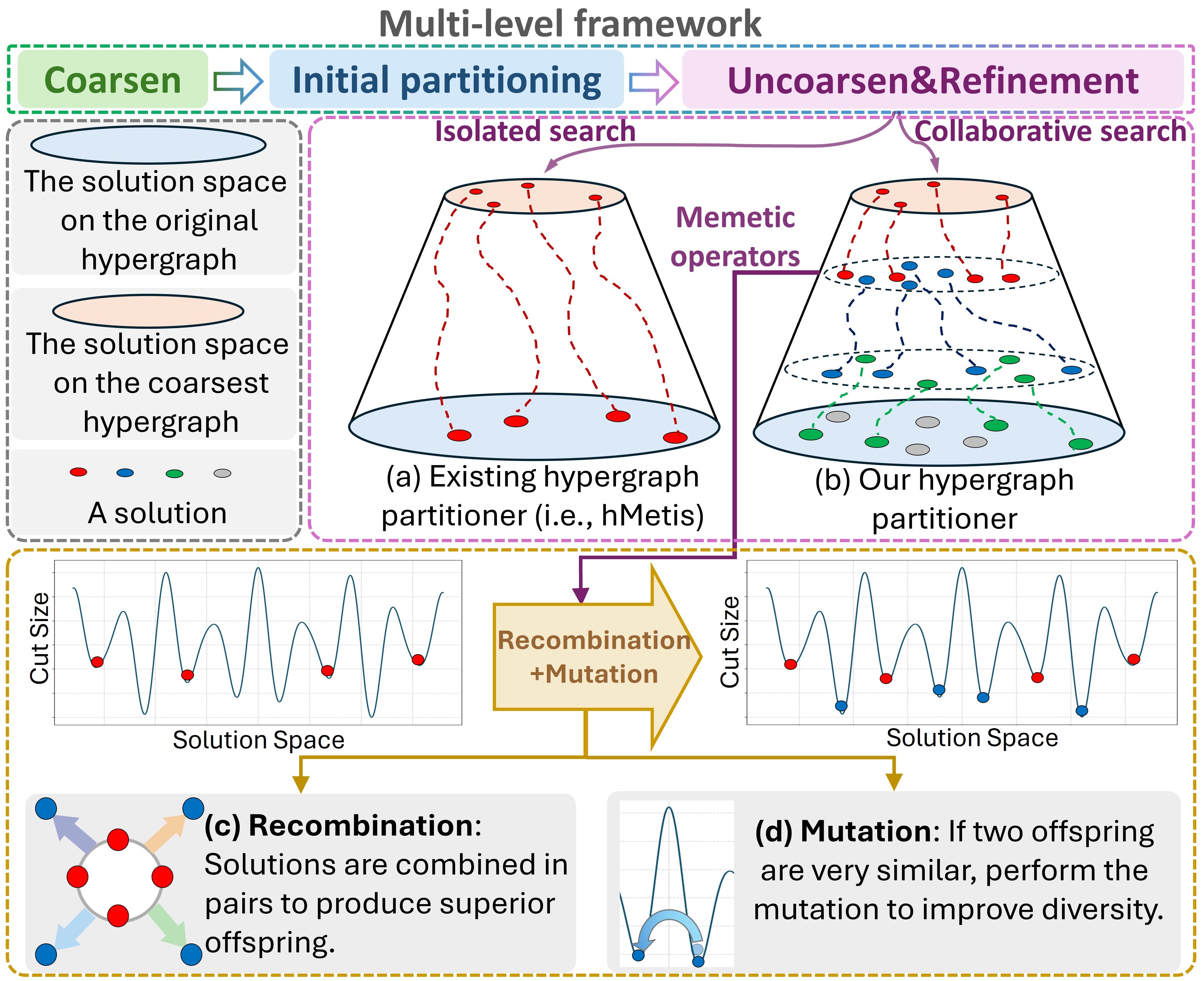}
    \vspace{-.2in}
    \caption{Framework comparison between the standard multi-level framework and ours.}
    \label{fig:framework}
    \vspace{-.15in}
\end{figure}

Most existing literature \cite{zhou2006learning,bustany2022specpart,gottesburen2019evaluation} has focused on the simpler case of bipartitioning ($k=2$).  Even for general $k$-way partitioning, many works \cite{bustany2023k,areibi2004effective,bustany2023open} only focus on relatively small cases, typically $k=2,3,4$. 
The general $k$-way partitioning problem, however, presents substantial challenges:
(1) Increasing the number of partitions $k$ leads to a combinatorial explosion of the solution space, as the number of possible ways to partition $n$ vertices is fundamentally governed by a $k^n$ term. This vast search landscape makes it increasingly difficult for heuristic algorithms to find a globally effective solution.
(2) Refinement becomes substantially more complex and local search heuristics such as FM are more prone to stalling in poor local optima.
Consequently, research into effective and scalable algorithms specifically tailored to the difficult $k$-way partitioning problem remains insufficient.

To address this gap, this paper \emph{introduces a new, open-source\footnote{https://github.com/hkust-zhiyao/IMPart} framework for the challenging large-$k$-way partitioning problem}, stemming from a fundamental overhaul of the conventional \emph{memetic algorithm} in hypergraph partitioning.

Memetic algorithm is an evolutionary method that evolves a population of solutions using operators such as recombination and mutation, and the individuals are improved via a local search. Although corresponding methods \cite{areibi2004effective,chan1993systolic,schwarz1999experimental} were applied to the hypergraph partitioning problem early, most of these initial approaches were rudimentary and non-multilevel. Consequently, such methods \cite{areibi2004effective,chan1993systolic,schwarz1999experimental} are not competitive with state-of-the-art multilevel hypergraph partitioners \cite{cohoon2003evolutionary,acikalin2022multilevel}. KaHyPar-E \cite{andre2018memetic} was the first to combine a memetic method with the multi-level partitioner, achieving results competitive with state-of-the-art partitioners and particularly demonstrating leading performance on large-$k$-way hypergraph partitioning problems. However, its approach was highly time-consuming, as \textbf{each recombination and mutation} operation relied on a \textbf{complete} multilevel partitioner to generate new solutions from existing ones.

In contrast, we propose \textbf{IMPart}, which thoroughly redesigns this memetic approach by integrating all recombination and mutation operations directly within a \textbf{single} multi-level partitioning process. This eliminates the need to repeatedly invoke complete external partitioners. This design is more efficient than the KaHyPar-E paradigm. Furthermore, IMPart introduces novel operators for recombination and mutation adapted to this integrated structure, establishing a new memetic paradigm.

Figure \ref{fig:framework} contrasts the traditional multi-level architecture (a) with our unified memetic framework (b). A fundamental architectural challenge exists in the traditional design: During uncoarsening, the local refinement is based on the partition structure inherited from the coarser level. Consequently, the multi-level framework is not flexible enough to adjust its partitioning scheme as the uncoarsening proceeds. The initial partition can exert an unpredictable influence on the final solution quality. Some methods, like hMETIS, attempt to improve solution quality by employing multiple initial partitions, as different initial solutions can explore different parts within the vast and unbounded solution space—as illustrated in Fig.\ref{fig:framework} (a). However, these independent searches operate in isolation, failing to leverage collective insights. Each run can still become trapped in its own local optimum, unable to learn from the partitioning schemes of others.

This challenge motivates our core insight: multiple solutions should explore the solution space collaboratively, providing mutual global guidance, as illustrated in Fig. \ref{fig:framework}(b). This population-based approach also lends itself to straightforward parallelization. To leverage the collaborative guidance, IMPart integrates novel memetic operators directly within the multi-level partitioning process:\looseness=-1

\textbf{\circled{1} Our recombination:} During uncoarsening, we periodically implement this collaborative learning via a recombination operator that merges the superior structures of different parent solutions. This approach is particularly effective for navigating the complex landscape of large-$k$-way hypergraph partitioning, allowing our framework to discover superior solutions. To manage these interactions and maintain a stable population, solutions are periodically reorganized into a ring topology (Fig. \ref{fig:framework}(c)), where each solution recombines with its neighbors. As these offspring are uncoarsened and refined, they develop new local structures, providing diverse material for future recombinations.

\textbf{\circled{2} Our mutation:} A drawback of the above recombination technique is that it may generate offspring that are too similar to each other. Since adjacent offspring in the ring-based structure share a common parent, they often inherit highly correlated features. Consequently, as the uncoarsening proceeds, offspring that have become overly similar tend to converge towards nearly identical partitioning schemes. This lack of diversity stifles the exploration of varied local structures from which solutions could mutually learn. To address this issue, we employ a diversity enhancement mechanism, as illustrated in Fig. \ref{fig:framework} (d). This mechanism identifies lower-quality solutions (i.e., those with a higher cut size) among similar ones, and replaces them with new offspring generated via a targeted mutation. This approach actively injects diversity into the population, guiding the search towards unexplored regions of the solution space.\looseness=-1

\begin{figure}[!t]
    \centering
    \includegraphics[width=\linewidth]{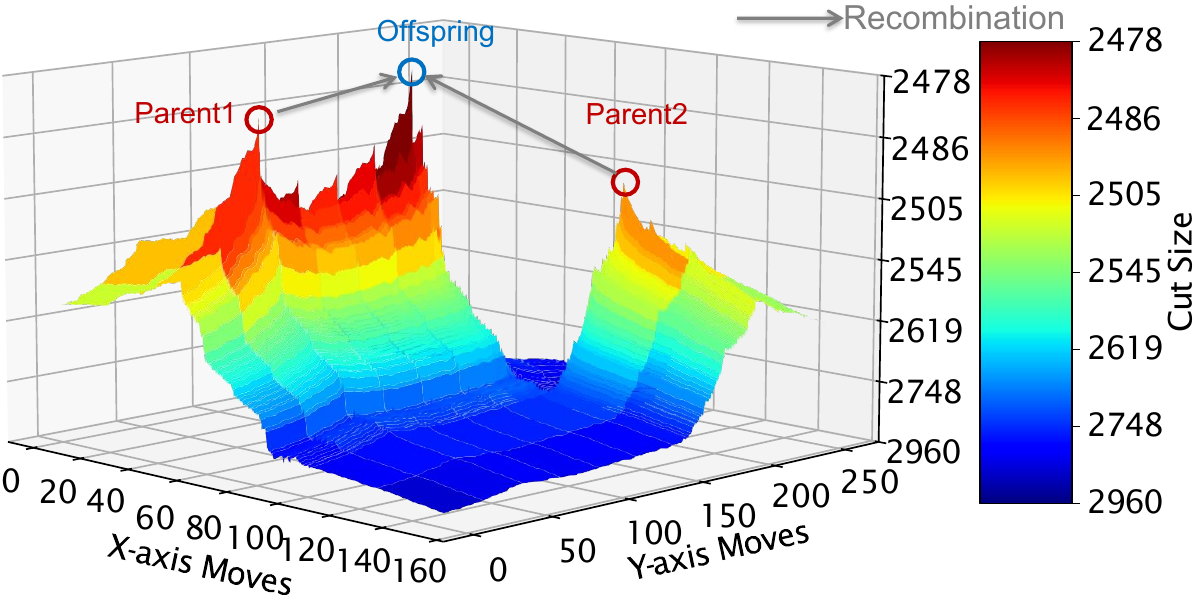}
    \vspace{-.1in}
    \caption{The jump mechanism in dataset \texttt{sparcT1\_core}: An improved global solution (offspring) is discovered through our recombination of parents.}
    \label{fig:case}
    \vspace{-.15in}
\end{figure}

These operators endow IMPart with a strong, periodic \emph{jumping} competence, as illustrated in Fig. \ref{fig:case}: parent solutions leap beyond local optima to produce a globally higher-quality offspring.

Evaluated on standard benchmarks (Titan23 \cite{murray2013titan}, ISPD98 \cite{alpert1998ispd98}), our algorithm consistently produced superior solutions compared to state-of-the-art (KaHyPar \cite{schlag2023high}), spectral (K-SpecPart \cite{bustany2023k}), foundational (hMETIS \cite{karypis1997multilevel}), and memetic (KaHyPar-E \cite{andre2018memetic}) methods. Our main contributions are:
\begin{itemize}
    \item \textbf{A memetics-integrated multi-level framework.} We propose IMPart, a novel memetic method embedded within a single multi-level partitioning process. This paradigm differs from existing memetic methods that repeatedly invoke the complete multi-level partitioner on the original hypergraph to perform each recombination and mutation. We designed novel recombination and mutation operators to support this structure. Our recombination during uncoarsening transforms the local searches of different granularities in the traditional multi-level framework into a sophisticated, collaborative search, enabling the discovery of globally higher-quality solutions, and inherently provides a parallelization opportunity to boost execution efficiency.\looseness=-1
    
    \item \textbf{Extensive experimental validation.} We provide both qualitative and quantitative evidence of our framework's effectiveness. We first conduct comprehensive comparisons against state-of-the-art partitioners on standard benchmarks, demonstrating that IMPart consistently achieves superior solution quality. Complementing these results, we visually demonstrate the distinct \emph{jumping} behavior that enables our method to effectively escape local optima, and finally, we verify the robust scalability of our algorithm in large-$k$ partitioning settings.
\end{itemize}

\section{Problem Formulation}
    \label{sec:pro}
    \begin{figure*}[!t]
    \centering
    \includegraphics[width=0.95\linewidth]{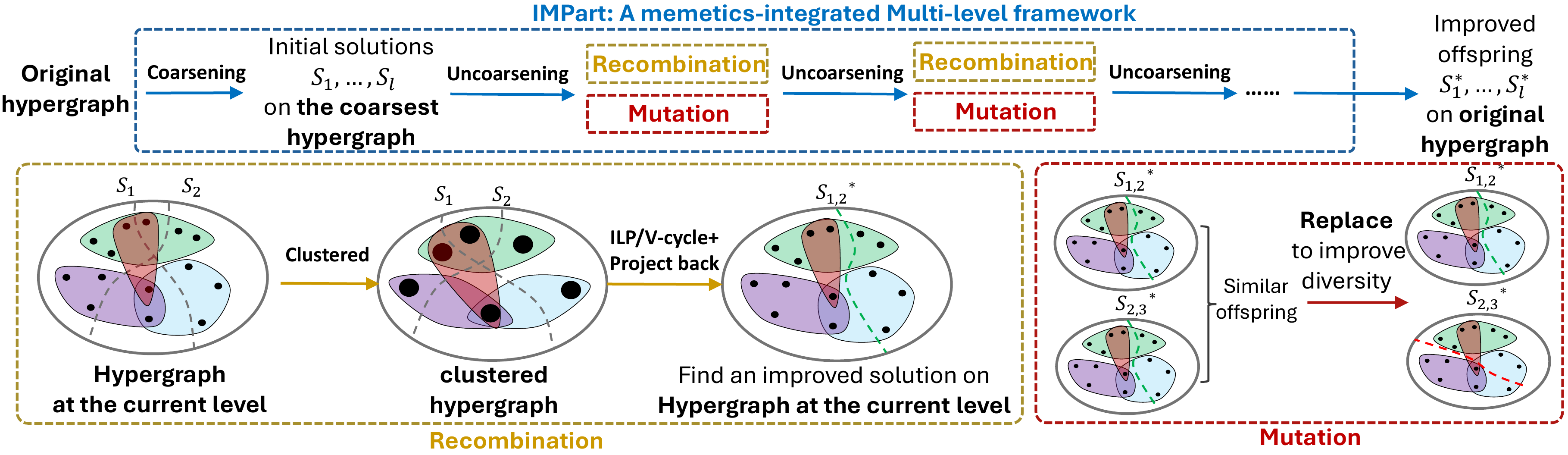}
    \vspace{-.15in}
    \caption{IMPart. Our memetics-integrated multi-level framework with recombination and mutation operators.}
    \label{fig:recombination}
    \vspace{-.15in}
\end{figure*}

Given a hypergraph $H(V,E)$ where $V$ is the set of vertices and $E$ is the set of hyperedges, the weights of all vertices and hyperedges are positive, associated with $w_v$ and $w_e$ respectively. The hypergraph partitioning problem\footnote{Our hypergraph partitioning formulation aligns with that of established tools (e.g., KaHyPar, PaToH, KaHyPar-E) and the recent hMETIS release (v2.0-pre1).} is to find a partition scheme $S$ that divides the nodes into $k$ disjoint blocks $V_1,V_2,...,V_k$ which $V_i\cap V_j=\emptyset$ and $ {\textstyle \bigcup_{i=0}^{k-1}} V_i=V$, the node weight $W_{V_i}$ of each block $i$ should satisfy: \looseness=-1
\begin{equation*}
    W_{V_i}\le(1+\epsilon)\cdot\left \lceil \frac{W_V}{k} \right \rceil    
\end{equation*}
\noindent where $W_V = \sum_{v \in V} w_v$, $W_{V_i} = \sum_{v \in V_i} w_v$ and $\epsilon$ is the imbalance factor. Under this constraint, the objective is to find the smallest $\text{cutsize}_H(S)$, where 
\begin{equation*}
    \text{cutsize}_H(S)=\sum_{e\not\subseteq V_i \text{ for all } V_i\in S}{w_e}
\end{equation*}

\section{Methodology}
    \label{sec:methodology}
    
Figure~\ref{fig:recombination} illustrates how our method, IMPart, enhances the standard multi-level paradigm by integrating two key evolutionary components: recombination and mutation. Recombination allows for mutual learning between different solutions, while mutation is employed to maintain population diversity and avoid premature convergence. Unlike the existing memetic method KaHyPar-E that repeatedly invokes the multi-level partitioner on the original hypergraph, IMPart integrates recombination and mutation \emph{internally}, making the framework more lightweight and efficient. It also elevates the local searches at different granularities during uncoarsening and refinement into a sophisticated, collaborative search.\looseness=-1

This section details the design and integration of these operators within a single multi-level partitioning process. We will first detail our recombination operator in Section \ref{sec:recombination}, and then describe the mutation operator in Section \ref{sec:mutation}. While our implementation builds upon the state-of-the-art KaHyPar framework, the proposed methods are generalizable to any multi-level partitioning approach.

\subsection{Recombination}
\label{sec:recombination}
Recombination is a technique for finding improved solutions from multiple existing partition schemes $S_1, S_2, \ldots, S_l$ for a hypergraph. We will first provide an overview of our recombination operator in Section \ref{sec:overview_of_re}, followed by its specific operational details in Section \ref{sec:operations_of_re}.

\subsubsection{Overview of Our Recombination}
\label{sec:overview_of_re}
Our recombination is \emph{embedded directly within the uncoarsening and refinement phase}. It is invoked periodically at different coarsening levels, allowing solutions to learn from one another and reposition their search to more globally advantageous regions of the solution space.

Specifically, our method first generates $\alpha$ diverse solutions (offspring). As uncoarsening progresses, these $\alpha$ solutions explore diverse directions in the solution space at each level, uncovering distinct partitioning schemes. Throughout the uncoarsening phase, we perform $\beta$ rounds of recombination at key moments ($\alpha=\beta=7$ in all our experiments). This process pairwise merges the current solutions to form $\alpha$ new, fused solutions in a circular manner\footnote{i.e., pairing offspring $1$ and $2$, $2$ and $3$, $\ldots$, $\alpha - 1$ and $\alpha$, and finally $\alpha$ and $1$.}. This allows each solution to break free from its original search trajectory and incorporate high-quality local structures from others. Then, as illustrated in Fig.~\ref{fig:framework}, the resulting offspring are superior solutions produced by recombination, which are positioned in more globally advantageous regions of the solution space. From there, they continue the search through subsequent uncoarsening and refinement. This architectural adjustment marks a significant departure from conventional methods. In existing multi-level frameworks, solutions in the uncoarsening phase are often restricted to minor improvements within the confines of the current local structure. Our proposed architecture, however, empowers these solutions to escape local optima, significantly enhancing their capacity to discover globally superior solutions.

The timing of these $\beta$ rounds is specifically tailored to the architecture of KaHyPar's $n$-level framework, where $n$ is the node number of the original hypergraph. In this paradigm, each coarsening step merges only a single pair of nodes, and refinement during uncoarsening is localized to the neighborhood of that node pair. We leverage its $n$-level framework to trigger recombination whenever the number of nodes being uncoarsened reaches a predefined threshold. Specifically, the set of thresholds is geometrically spaced over the uncoarsening trajectory as follows:
\[
\left\{ n_c^{1 - 1/\beta} n^{1/\beta},\; n_c^{1 - 2/\beta} n^{2/\beta},\; \ldots,\; n_c^{1/\beta} n^{1 - 1/\beta},\; n \right\},
\]
where $n_c$ is the node number of the coarsest hypergraph. 

We will now describe the detailed operation of recombination, explaining how to generate a globally higher-quality solution from two given solutions at a specific coarsening level.

\subsubsection{The Specific Operations of Our Recombination}
\label{sec:operations_of_re}
 We aim to produce superior solutions by leveraging a recombination process that integrates beneficial features from two parent solutions. The methodology involves first applying recombination to hypergraphs at various levels to yield a clustered hypergraph as illustrated in Fig.~\ref{fig:recombination}. Subsequently, we employ a hybrid strategy to process the clustered hypergraph based on its complexity. For instances below a predefined threshold, we apply an integer linear programming (ILP) formulation \cite{bustany2022specpart,henzinger2020ilp}. While ILP guarantees an optimal solution, its significant time complexity is prohibitive for larger cases. Therefore, for hypergraphs exceeding this threshold, we resort to the V-cycle in KaHyPar, a technique applied to improve the current solution.

The constraints in ILP are defined as below. We introduce integer $\{0,1\}$ variables: $x_{v,i}$ for each vertex $v$ and for each block $V_i$, and $y_{e,i}$ for each hyperedge $e$ and for each block $V_i$. Specifically, $x_{v,i}=1$ signifies that vertex $v$ is in block $V_i$. The variable $y_{e,i} = 1$ specifically means all vertices of hyperedge $e$ are entirely contained within block $V_i$. Then we can define the hypergraph partitioning problem as below:

\begin{itemize}
    \item $\sum_{j=0}^{K-1} x_{v,j} = 1$, for all $v \in V$
    \item $y_{e,i} \leq x_{v,i}$ for all $e \in E$, and $v \in e$
    \item $\sum_{v \in V_i} w_v x_{v,i} \leq \left(1 + \epsilon\right)\cdot\left \lceil \frac{W_V}{k} \right \rceil $ \\
    
\end{itemize}
The objective is to maximize the total weight of the hyperedges that are not cut, i.e.,
\[
\text{maximize } \sum_{e \in E} \sum_{i=0}^{K-1} w_e y_{e,i}.
\]

Specifically, we use C++ Gurobi to tackle the aforementioned ILP problem. Since the hypergraph partitioning problem is NP-hard, its solution space is vast. Therefore, we employ several strategies to accelerate the ILP solver:

\begin{enumerate}
    \item \textbf{Symmetry Breaking.} To eliminate redundant symmetric solutions (i.e., block ID permutations), we restrict vertex $v$ to be assigned only to partitions $i$ such that $i \le v$.
    \item \textbf{Warm Start.} To accelerate convergence, we initialize the ILP solver with the better of the two parent solutions (the one with the smaller cut size).
\end{enumerate}

We let $n'$ be the number of nodes in the clustered hypergraph and use the metric $n' \times k$ to determine the solution method. For metrics below 600, we use the ILP solver to find a provably optimal solution. For metrics in the range $[600, 1000)$, we configure the ILP for an approximate solution with an optimality gap of at most 1\%. Finally, for metrics $\ge 1000$, we resort to the V-cycle method.

\subsection{Mutation}
    \label{sec:mutation}
While the technique discussed in the previous subsection is effective, a common challenge arises as uncoarsening progresses: due to mutual learning, offspring often converge to similar or even identical structures. To counteract this loss of diversity and enable a broader exploration of the solution space, we employ a mutation operator to ensure differentiation among offspring.

\textbf{The mutation operators.} After executing the recombination operator, we sort all the obtained new offspring by their cut size in ascending order and process them sequentially. For each selected offspring $S_i$, we calculate its similarity to every subsequent offspring $S_j(j>i)$ using an edge-based distance $d_e(S_i,S_j)$. If this metric is less than a predefined threshold $t$ ($t=20$ in our experiments), offspring $S_i$ is added to the set $M(S_j)$. If an offspring $S_j$ has an empty set $M(S_j)$ (meaning it's sufficiently distinct from all preceding offspring), it does not undergo mutation. Otherwise, it is mutated based on the offspring within its $M(S_j)$ set. 

The guiding principle is to mutate $S_j$ in a way that not only seeks a small cut-size but also increases its distance from the solutions in $M(S_j)$. To achieve this, we construct a new hypergraph where the sets of vertices and hyperedges are identical to the current-level hypergraph, and vertex weights remain unchanged. However, the weights of the hyperedges are updated as $w_e' = w_e (1 + 0.1 \cdot C_{M(S_j)}(e))$, where $C_{M(S_j)}(e)$ represents the count of offspring in $M(S_j)$ that cut hyperedge $e$. We then process this new hypergraph using the partitioner KaHyPar, and the resulting partition scheme replaces the original solution $S_j$. The more times a hyperedge is cut in $M(S_j)$, the greater its weight is increased, and the lower the likelihood that the hypergraph partitioner will cut it. Consequently, the updated solution $S_j$ is guided to be structurally dissimilar to the offspring in $M(S_j)$, which helps maintain population diversity at this level.

\begin{figure}[t]
    \centering
    \includegraphics[width=0.85\linewidth]{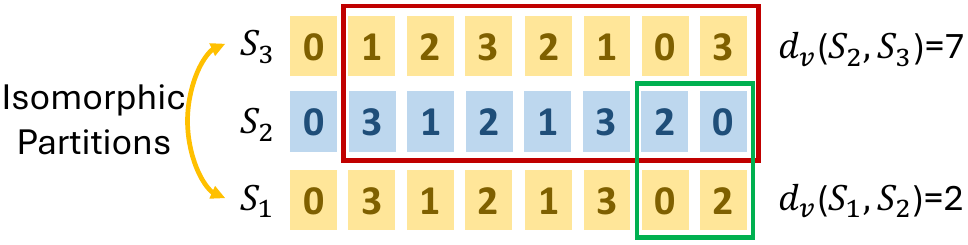}
    \caption{An example of the partition isomorphism problem in node-based metric.}
    \label{fig:nodepart}
    \vspace{-.15in}
\end{figure}

\textbf{The measurement of similarity.} To measure the similarity between two partitions, $S_i$ and $S_j$, an intuitive \textbf{node-based metric} is the Hamming distance, which counts the number of mismatched node assignments:
\begin{equation}
   d_v(S_i, S_j) = \sum_{k=1}^{|V|} \delta_k, \quad \text{where } \delta_k = 
   \begin{cases}
       1 & \text{if } S_i(k) \neq S_j(k) \\
       0 & \text{if } S_i(k) = S_j(k)
   \end{cases}
   \label{eq:node_dist}
\end{equation}
\noindent where $S_i(k)$ denotes the partition ID of node $k$ in partition $S_i$. However, this metric is susceptible to the \textbf{partition isomorphism problem}, where identical partitions can be measured as different simply due to a relabeling of the partition groups (an example is shown in Fig.~\ref{fig:nodepart}).

\begin{table*}[!htbp]
\centering

\caption{Experimental results on Titan23 benchmark.}
\vspace{-.1in}
\label{tab:Titan23}
\begingroup 
\setlength{\tabcolsep}{2.5pt}
\resizebox{0.99\textwidth}{!}{

\begin{NiceTabular}{c||c|c|c|c|c||c|c|c|c||c|c|c|c||c|c|c|c} \toprule
\multirow{3}{*}{ Design }       & \multicolumn{5}{c||}{$k=4, \text{imbalance}^*=2\%$} &
       \multicolumn{4}{c||}{$k=4, \text{imbalance}=5\%$} &
       \multicolumn{4}{c||}{$k=10, \text{imbalance}=2\%$} &
       \multicolumn{4}{c}{$k=10, \text{imbalance}=5\%$}
       \\ \cline{2-18}
\textbf{\begin{tabular}[c]{@{}c@{}} \phantom{1} \\ \phantom{2}\end{tabular}}&  \hspace{-3pt} \cellcolor[HTML]{F2F2F2}\textbf{hMETIS} \hspace{-3pt} &\cellcolor[HTML]{F2F2F2}\textbf{K-Spec}  & \hspace{1pt}\cellcolor[HTML]{F2F2F2}\textbf{KaHy}\hspace{1pt} & \hspace{-2pt}\cellcolor[HTML]{F2F2F2}\textbf{KaHy-E} \hspace{-2pt}  & \hspace{1pt} \cellcolor[HTML]{FFF2CC}\textbf{Ours} \hspace{1pt}   & \hspace{-3pt} \cellcolor[HTML]{F2F2F2}\textbf{hMETIS} \hspace{-3pt}    & \hspace{1pt} \cellcolor[HTML]{F2F2F2}\textbf{KaHy} \hspace{1pt} & \cellcolor[HTML]{F2F2F2}\textbf{KaHy-E}   & \hspace{1pt} \cellcolor[HTML]{FFF2CC}\textbf{Ours} \hspace{1pt}  & \hspace{-3pt} \cellcolor[HTML]{F2F2F2}\textbf{hMETIS}  \hspace{-3pt}    & \hspace{1pt} \cellcolor[HTML]{F2F2F2}\textbf{KaHy} \hspace{1pt} & \cellcolor[HTML]{F2F2F2}\textbf{KaHy-E}  & \hspace{1pt} \cellcolor[HTML]{FFF2CC}\textbf{Ours} \hspace{1pt}  & \hspace{-3pt} \cellcolor[HTML]{F2F2F2}\textbf{hMETIS} \hspace{-3pt}  & \hspace{1pt} \cellcolor[HTML]{F2F2F2}\textbf{KaHy} \hspace{1pt} & \cellcolor[HTML]{F2F2F2}\textbf{KaHy-E} & \hspace{1pt} \cellcolor[HTML]{FFF2CC}\textbf{Ours} \hspace{1pt} \\
\hline \hline
sparcT1\_core  & 2633                                 & 2492 & 2646                                & 2485                                 & {\color[HTML]{00B050} \textbf{2229}} & 2573 & 2467                                 & {\color[HTML]{00B050} \textbf{2313}} & 2315                                 & 3790                                 & 3721                                 & 3682                                 & {\color[HTML]{00B050} \textbf{3596}} & 3884 & 3452                                 & 3524                                 & {\color[HTML]{00B050} \textbf{3291}} \\
neuron        & 580                                  & 431  & 422                                 & 424                                  & {\color[HTML]{00B050} \textbf{405}}  & 505  & 415                                  & 410                                  & {\color[HTML]{00B050} \textbf{406}}  & 1032                                 & 724                                  & 728                                  & {\color[HTML]{00B050} \textbf{684}}  & 1058 & 659                                  & 652                                  & {\color[HTML]{00B050} \textbf{647}}  \\
stereo\_vision & 455                                  & 475  & 394                                 & {\color[HTML]{00B050} \textbf{367}}  & {\color[HTML]{00B050} \textbf{367}}  & 424  & {\color[HTML]{00B050} \textbf{315}}  & 317                                  & {\color[HTML]{00B050} \textbf{315}}  & 773                                  & 732                                  & {\color[HTML]{00B050} \textbf{724}}  & 728                                  & 730  & 681                                  & {\color[HTML]{00B050} \textbf{652}}  & 678                                  \\
des90         & 700                                  & 747  & 699                                 & {\color[HTML]{00B050} \textbf{685}}  & {\color[HTML]{00B050} \textbf{685}}  & 768  & 625                                  & {\color[HTML]{00B050} \textbf{623}}  & 628                                  & 1494                                 & 1301                                 & 1309                                 & {\color[HTML]{00B050} \textbf{1252}} & 1543 & 1158                                 & 1128                                 & {\color[HTML]{00B050} \textbf{1036}} \\
SLAM\_spheric  & 3423                                 & 3241 & 3217                                & 3201                                 & {\color[HTML]{00B050} \textbf{3191}} & 3354 & 3114                                 & {\color[HTML]{00B050} \textbf{3056}} & 3103                                 & 5029                                 & 4456                                 & 4361                                 & {\color[HTML]{00B050} \textbf{4305}} & 4931 & 4400                                 & 4205                                 & {\color[HTML]{00B050} \textbf{4102}} \\
cholesky\_mc   & 983                                  & 984  & {\color[HTML]{00B050} \textbf{975}} & {\color[HTML]{00B050} \textbf{975}}  & {\color[HTML]{00B050} \textbf{975}}  & 984  & {\color[HTML]{00B050} \textbf{965}}  & {\color[HTML]{00B050} \textbf{965}}  & {\color[HTML]{00B050} \textbf{965}}  & 2229                                 & 1836                                 & {\color[HTML]{00B050} \textbf{1773}} & 1777                                 & 2121 & 1700                                 & 1673                                 & {\color[HTML]{00B050} \textbf{1667}} \\
segmentation  & 557                                  & 490  & 478                                 & {\color[HTML]{00B050} \textbf{477}}  & {\color[HTML]{00B050} \textbf{477}}  & 531  & {\color[HTML]{00B050} \textbf{443}}  & {\color[HTML]{00B050} \textbf{443}}  & {\color[HTML]{00B050} \textbf{443}}  & 1536                                 & 1238                                 & 1139                                 & {\color[HTML]{00B050} \textbf{1115}} & 1498 & 883                                  & {\color[HTML]{00B050} \textbf{879}}  & {\color[HTML]{00B050} \textbf{879}}  \\
bitonic\_mesh  & 1113                                 & 1311 & 1100                                & 1094                                 & {\color[HTML]{00B050} \textbf{1091}} & 1115 & 1088                                 & 1086                                 & {\color[HTML]{00B050} \textbf{1084}} & 2388                                 & 2149                                 & 2212                                 & {\color[HTML]{00B050} \textbf{2102}} & 2340 & 1962                                 & 1929                                 & {\color[HTML]{00B050} \textbf{1894}} \\
dart          & 1449                                 & 1401 & 1302                                & 1253                                 & {\color[HTML]{00B050} \textbf{1242}} & 1475 & 1027                                 & {\color[HTML]{00B050} \textbf{1025}} & {\color[HTML]{00B050} \textbf{1025}} & 1663                                 & 1529                                 & 1528                                 & {\color[HTML]{00B050} \textbf{1527}} & 1693 & 1630                                 & 1528                                 & {\color[HTML]{00B050} \textbf{1527}} \\
openCV        & 574                                  & 522  & 670                                 & 519                                  & {\color[HTML]{00B050} \textbf{514}}  & 597  & 575                                  & 531                                  & {\color[HTML]{00B050} \textbf{497}}  & 970                                  & 727                                  & 713                                  & {\color[HTML]{00B050} \textbf{681}}  & 1052 & 752                                  & 695                                  & {\color[HTML]{00B050} \textbf{645}}  \\
stap\_qrd      & 712                                  & 674  & 612                                 & {\color[HTML]{00B050} \textbf{611}}  & 614                                  & 710  & 612                                  & 547                                  & {\color[HTML]{00B050} \textbf{539}}  & 1472                                 & 972                                  & 972                                  & {\color[HTML]{00B050} \textbf{967}}  & 1410 & 881                                  & 878                                  & {\color[HTML]{00B050} \textbf{875}}  \\
minres        & 411                                  & 407  & 411                                 & {\color[HTML]{00B050} \textbf{405}}  & 411                                  & 413  & 347                                  & 347                                  & {\color[HTML]{00B050} \textbf{341}}  & 866                                  & 799                                  & 765                                  & {\color[HTML]{00B050} \textbf{764}}  & 882  & 699                                  & {\color[HTML]{00B050} \textbf{662}}  & 667                                  \\
cholesky\_bdti & 1901                                 & 1865 & 1913                                & 1863                                 & {\color[HTML]{00B050} \textbf{1861}} & 1891 & 1793                                 & {\color[HTML]{00B050} \textbf{1791}} & {\color[HTML]{00B050} \textbf{1791}} & 3785                                 & 3154                                 & {\color[HTML]{00B050} \textbf{3152}} & {\color[HTML]{00B050} \textbf{3152}} & 3897 & 3172                                 & 3140                                 & {\color[HTML]{00B050} \textbf{3126}} \\
denoise       & 953                                  & 1001 & 944                                 & 872                                  & {\color[HTML]{00B050} \textbf{861}}  & 1166 & 870                                  & 865                                  & {\color[HTML]{00B050} \textbf{814}}  & 1906                                 & 1501                                 & 1496                                 & {\color[HTML]{00B050} \textbf{1487}} & 1889 & 1237                                 & {\color[HTML]{00B050} \textbf{1116}} & {\color[HTML]{00B050} \textbf{1116}} \\
sparcT2\_core  & 3165                                 & 3558 & 2682                                & 2696                                 & {\color[HTML]{00B050} \textbf{2680}} & 2506 & 2415                                 & 2374                                 & {\color[HTML]{00B050} \textbf{2350}} & 5220                                 & 5125                                 & 4945                                 & {\color[HTML]{00B050} \textbf{4887}} & 5438 & 3826                                 & 3817                                 & {\color[HTML]{00B050} \textbf{3779}} \\
gsm\_switch    & 5354                                 & 4404 & 2883                                & 2901                                 & {\color[HTML]{00B050} \textbf{2841}} & 5289 & 2692                                 & 2692                                 & {\color[HTML]{00B050} \textbf{2681}} & 6056                                 & 6139                                 & 6139                                 & {\color[HTML]{00B050} \textbf{5820}} & 5923 & 5078                                 & 5377                                 & {\color[HTML]{00B050} \textbf{4950}} \\
mes\_noc       & 1400                                 & 1346 & 1332                                & {\color[HTML]{00B050} \textbf{1307}} & {\color[HTML]{00B050} \textbf{1307}} & 1406 & 1262                                 & 1262                                 & {\color[HTML]{00B050} \textbf{1255}} & 3251                                 & 2616                                 & 2616                                 & {\color[HTML]{00B050} \textbf{2581}} & 3156 & 2493                                 & 2353                                 & {\color[HTML]{00B050} \textbf{2296}} \\
LU230         & {\color[HTML]{00B050} \textbf{5794}} & 6310 & 5821                                & 6001                                 & 5905                                 & 5755 & 5889                                 & 5938                                 & {\color[HTML]{00B050} \textbf{5591}} & {\color[HTML]{00B050} \textbf{8997}} & 9568                                 & 9568                                 & 9290                                 & 9046 & 9185                                 & 9048                                 & {\color[HTML]{00B050} \textbf{8924}} \\
LU\_Network    & 1509                                 & 1417 & 1247                                & 1252                                 & {\color[HTML]{00B050} \textbf{1246}} & 1466 & {\color[HTML]{00B050} \textbf{1044}} & {\color[HTML]{00B050} \textbf{1044}} & {\color[HTML]{00B050} \textbf{1044}} & 3988                                 & 3446                                 & {\color[HTML]{00B050} \textbf{3442}} & {\color[HTML]{00B050} \textbf{3442}} & 3918 & 2086                                 & 2086                                 & {\color[HTML]{00B050} \textbf{2084}} \\
sparcT1\_chip2 & 1759                                 & 1601 & 1431                                & 1602                                 & {\color[HTML]{00B050} \textbf{1399}} & 1773 & {\color[HTML]{00B050} \textbf{1216}} & {\color[HTML]{00B050} \textbf{1218}} & {\color[HTML]{00B050} \textbf{1216}} & 3900                                 & {\color[HTML]{00B050} \textbf{2688}} & {\color[HTML]{00B050} \textbf{2688}} & 2778                                 & 3410 & 2236                                 & 2467                                 & {\color[HTML]{00B050} \textbf{2159}} \\
directrf      & 1070                                 & 1092 & 1069                                & 1209                                 & {\color[HTML]{00B050} \textbf{972}}  & 1065 & 955                                  & 959                                  & {\color[HTML]{00B050} \textbf{933}}  & 2594                                 & 2534                                 & 2528                                 & {\color[HTML]{00B050} \textbf{2522}} & 2746 & {\color[HTML]{00B050} \textbf{2081}} & 2104                                 & 2104                                 \\
bitcoin\_miner & 2601                                 & 2737 & 1862                                & 1896                                 & {\color[HTML]{00B050} \textbf{1612}} & 2269 & 1349                                 & 1382                                 & {\color[HTML]{00B050} \textbf{1294}} & 5496                                 & 3142                                 & 2984                                 & {\color[HTML]{00B050} \textbf{2926}} & 5108 & 1701                                 & {\color[HTML]{00B050} \textbf{1699}} & 1700    \\  \hline
\textbf{\begin{tabular}[c]{@{}c@{}} Norm \\ Avg.\end{tabular}}    &  \cellcolor[HTML]{F2F2F2}1.113  & \cellcolor[HTML]{F2F2F2}1.086       & \cellcolor[HTML]{F2F2F2}1 & \cellcolor[HTML]{F2F2F2}0.987   & \cellcolor[HTML]{FFF2CC}\textbf{0.954} & \cellcolor[HTML]{F2F2F2}1.213  & \cellcolor[HTML]{F2F2F2}1& \cellcolor[HTML]{F2F2F2}0.988  & \cellcolor[HTML]{FFF2CC}\textbf{0.973}     & \cellcolor[HTML]{F2F2F2}1.186& \cellcolor[HTML]{F2F2F2}1        & \cellcolor[HTML]{F2F2F2}0.988    & \cellcolor[HTML]{FFF2CC}\textbf{0.971}      & \cellcolor[HTML]{F2F2F2}1.365& \cellcolor[HTML]{F2F2F2}1& \cellcolor[HTML]{F2F2F2}0.983 & \cellcolor[HTML]{FFF2CC}\textbf{0.961}   \\ \bottomrule

\end{NiceTabular}
}
\endgroup
\par\vspace{2pt}
{\raggedright \footnotesize
$^\star$ The imbalance is set as a percentage ($p$) of the total number of vertices. It relates to the imbalance factor $\epsilon$ via $\epsilon = k \cdot p$.
\par}
\vspace{-.1in}
\end{table*}

Therefore, we adopt a more robust, label-invariant \textbf{edge-based metric}. This metric is the $L_1$ distance between the connectivity vectors of the two partitions:
\begin{equation}
   d_e(S_i, S_j) = \sum_{e\in E}{\lvert\mathrm{connectivity}_{S_i}(e)-\mathrm{connectivity}_{S_j}(e)\rvert}
   \label{eq:edge_dist}
\end{equation}
\noindent where $\mathrm{connectivity}_{S_i}(e)$ is the number of distinct partitions spanned by hyperedge $e$ in partition $S_i$. This metric is therefore well-aligned with the primary goal of hypergraph partitioning, as it directly quantifies differences in the cut structure.

\section{experiments}
    \label{sec:exp}
    In this section, we first introduce the experimental setup in Section~\ref{exp:setup}. Then, in Section~\ref{exp:comparison}, we conduct comprehensive performance comparisons on standard benchmarks including Titan23 \cite{murray2013titan} and ISPD98 \cite{alpert1998ispd98}.

\subsection{Experimental Setup}
\label{exp:setup}

Our experimental evaluation comprises three main components:
\begin{enumerate}
    \item Sections~\ref{exp:titan23} and~\ref{exp:ispd98} present overall comparisons between our method and leading hypergraph partitioners, including hMETIS, K-SpecPart (K-Spec), KaHyPar (KaHy), and KaHyPar-E (KaHy-E), on the Titan23 and ISPD98 benchmarks, respectively. We focus on the multi-way partitioning problem, with experiments conducted for $k=4$ and $k=10$ partitions. We establish two imbalance constraints by setting the maximum allowable size to the average size ($|V|/k$) plus an additional 2\% or 5\% of the total node count ($|V|$).\footnote{For $k=4$ partitions, these constraints translate directly to the standard imbalance factors ($\epsilon$) of 0.08 and 0.20. For $k=10$ partitions, they correspond to $\epsilon$ values of 0.20 and 0.50, respectively.} We provide a normalized average (Norm. Avg.) for each method, calculated as the geometric mean relative to KaHyPar. To ensure a fair comparison, our method (IMPart), hMETIS, and KaHyPar were tested against \textbf{the same total execution time}, and the best result obtained within that time was reported. Due to library compatibility issues (missing \texttt{triton\_part\_refine} command) preventing the execution of K-SpecPart, we can only report the results for K-SpecPart ($k=4$, imbalance$=2\% $) directly from its original publication. Furthermore, owing to its inherent complexity, KaHyPar-E was allocated double this execution time.
    \item In Section \ref{exp:jump}, we visualize the optimization process to provide evidence of our framework's ability to escape local optima. We contrast our method's trajectory, characterized by abrupt, significant improvements, with baseline approaches that tend to show only gradual refinement.
    \item In Section~\ref{exp:scalability}, we extend our analysis to larger $k$ values ($k=16$ and $32$), presenting an overall comparison of cut size performance against the baselines to demonstrate the scalability of our algorithm.
\end{enumerate} 

We have open-sourced all our code and released all partitioning results for reproducibility. Our implementation is based on KaHyPar, with all parameters consistent with it\footnote{The configuration document is \emph{cut\_kKaHyPar\_sea20.ini}.}. All experiments were conducted on a \texttt{single core} of an \texttt{Intel(R) Xeon(R) Gold 6438Y+ CPU @ 2.00GHz}, running \texttt{Ubuntu 22.04.5 LTS} with \texttt{512GB RAM}. Our implementation was compiled from source using \texttt{g++ 11.4.0} with the \texttt{C++17} standard.

\begin{table*}[t]
\centering
\caption{Experimental results on ISPD98 benchmark.}
\vspace{-.1in}
\label{tab:ISPD98}
\begingroup 
\setlength{\tabcolsep}{2.5pt}
\resizebox{0.99\textwidth}{!}{

\begin{NiceTabular}{c||c|c|c|c|c||c|c|c|c||c|c|c|c||c|c|c|c} \toprule
\multirow{3}{*}{ Design }       & \multicolumn{5}{c||}{$k=4, \text{imbalance}=2\%$} &
       \multicolumn{4}{c||}{$k=4, \text{imbalance}=5\%$} &
       \multicolumn{4}{c||}{$k=10, \text{imbalance}=2\%$} &
       \multicolumn{4}{c}{$k=10, \text{imbalance}=5\%$}
       \\ \cline{2-18}
\textbf{\begin{tabular}[c]{@{}c@{}} \phantom{1} \\ \phantom{2}\end{tabular}} &\hspace{-3pt} \cellcolor[HTML]{F2F2F2}\textbf{hMETIS} \hspace{-3pt} &\cellcolor[HTML]{F2F2F2}\textbf{K-Spec}  & \hspace{1pt}\cellcolor[HTML]{F2F2F2}\textbf{KaHy}\hspace{1pt} & \hspace{-2pt}\cellcolor[HTML]{F2F2F2}\textbf{KaHy-E} \hspace{-2pt}  & \hspace{1pt} \cellcolor[HTML]{FFF2CC}\textbf{Ours} \hspace{1pt}   & \hspace{-3pt} \cellcolor[HTML]{F2F2F2}\textbf{hMETIS} \hspace{-3pt}    & \hspace{1pt} \cellcolor[HTML]{F2F2F2}\textbf{KaHy} \hspace{1pt} & \cellcolor[HTML]{F2F2F2}\textbf{KaHy-E}   & \hspace{1pt} \cellcolor[HTML]{FFF2CC}\textbf{Ours} \hspace{1pt}  & \hspace{-3pt} \cellcolor[HTML]{F2F2F2}\textbf{hMETIS}  \hspace{-3pt}    & \hspace{1pt} \cellcolor[HTML]{F2F2F2}\textbf{KaHy} \hspace{1pt} & \cellcolor[HTML]{F2F2F2}\textbf{KaHy-E}  & \hspace{1pt} \cellcolor[HTML]{FFF2CC}\textbf{Ours} \hspace{1pt}  & \hspace{-3pt} \cellcolor[HTML]{F2F2F2}\textbf{hMETIS} \hspace{-3pt}  & \hspace{1pt} \cellcolor[HTML]{F2F2F2}\textbf{KaHy} \hspace{1pt} & \cellcolor[HTML]{F2F2F2}\textbf{KaHy-E} & \hspace{1pt} \cellcolor[HTML]{FFF2CC}\textbf{Ours} \hspace{1pt} \\
\hline \hline
ibm01 & 476  & 522  & 462                                  & 458                                  & {\color[HTML]{00B050} \textbf{450}}  & 458  & 416                                  & 416                                  & {\color[HTML]{00B050} \textbf{415}}  & 880   & 808                                  & 807                                  & {\color[HTML]{00B050} \textbf{796}}  & 887   & 699  & 697                                  & {\color[HTML]{00B050} \textbf{695}}  \\
ibm02 & 596  & 706  & {\color[HTML]{00B050} \textbf{584}}  & 618                                  & 607                                  & 596  & 553                                  & 550                                  & {\color[HTML]{00B050} \textbf{528}}  & 1978  & 1896                                 & 1893                                 & {\color[HTML]{00B050} \textbf{1888}} & 1832  & 1602 & 1595                                 & {\color[HTML]{00B050} \textbf{1589}} \\
ibm03 & 1672 & 1690 & 1659                                 & 1647                                 & {\color[HTML]{00B050} \textbf{1625}} & 1638 & 1647                                 & {\color[HTML]{00B050} \textbf{1646}} & {\color[HTML]{00B050} \textbf{1646}} & 2659  & 2617                                 & 2560                                 & {\color[HTML]{00B050} \textbf{2510}} & 2643  & 2518 & 2428                                 & {\color[HTML]{00B050} \textbf{2391}} \\
ibm04 & 1620 & 1626 & {\color[HTML]{00B050} \textbf{1590}} & {\color[HTML]{00B050} \textbf{1590}} & {\color[HTML]{00B050} \textbf{1590}} & 1636 & 1524                                 & 1517                                 & {\color[HTML]{00B050} \textbf{1512}} & 3203  & 3007                                 & 2921                                 & {\color[HTML]{00B050} \textbf{2827}} & 3103  & 2615 & 2539                                 & {\color[HTML]{00B050} \textbf{2531}} \\
ibm05 & 2946 & 2946 & 2960                                 & {\color[HTML]{00B050} \textbf{2897}} & 2945                                 & 2913 & 2901                                 & 2878                                 & {\color[HTML]{00B050} \textbf{2855}} & 4656  & 4205                                 & 4117                                 & {\color[HTML]{00B050} \textbf{4106}} & 4451  & 4005 & 3827                                 & {\color[HTML]{00B050} \textbf{3820}} \\
ibm06 & 1481 & 1476 & 1467                                 & {\color[HTML]{00B050} \textbf{1465}} & {\color[HTML]{00B050} \textbf{1465}} & 1486 & 1466                                 & 1461                                 & {\color[HTML]{00B050} \textbf{1455}} & 2662  & 2531                                 & 2454                                 & {\color[HTML]{00B050} \textbf{2417}} & 2569  & 2261 & {\color[HTML]{00B050} \textbf{2187}} & 2232                                 \\
ibm07 & 2176 & 2154 & 2068                                 & 2049                                 & {\color[HTML]{00B050} \textbf{2045}} & 2148 & 1978                                 & {\color[HTML]{00B050} \textbf{1954}} & {\color[HTML]{00B050} \textbf{1954}} & 3758  & 3465                                 & 3436                                 & {\color[HTML]{00B050} \textbf{3392}} & 3730  & 3140 & 3081                                 & {\color[HTML]{00B050} \textbf{3033}} \\
ibm08 & 2331 & 2328 & 2292                                 & {\color[HTML]{00B050} \textbf{2278}} & 2286                                 & 2313 & 2185                                 & {\color[HTML]{00B050} \textbf{2178}} & 2188                                 & 3774  & 3521                                 & 3441                                 & {\color[HTML]{00B050} \textbf{3419}} & 3727  & 3299 & {\color[HTML]{00B050} \textbf{3178}} & 3212                                 \\
ibm09 & 1700 & 1676 & 1672                                 & 1672                                 & {\color[HTML]{00B050} \textbf{1651}} & 1711 & 1594                                 & {\color[HTML]{00B050} \textbf{1583}} & 1586                                 & 3187  & 2768                                 & 2672                                 & {\color[HTML]{00B050} \textbf{2665}} & 3223  & 2557 & 2563                                 & {\color[HTML]{00B050} \textbf{2462}} \\
ibm10 & 2433 & 2400 & 2190                                 & {\color[HTML]{00B050} \textbf{2188}} & {\color[HTML]{00B050} \textbf{2188}} & 2396 & {\color[HTML]{00B050} \textbf{2155}} & {\color[HTML]{00B050} \textbf{2155}} & 2156                                 & 4599  & 4395                                 & 4128                                 & {\color[HTML]{00B050} \textbf{4122}} & 4537  & 3669 & 3555                                 & {\color[HTML]{00B050} \textbf{3522}} \\
ibm11 & 2476 & 2452 & 2346                                 & 2341                                 & {\color[HTML]{00B050} \textbf{2327}} & 2459 & 2035                                 & 2033                                 & {\color[HTML]{00B050} \textbf{2032}} & 4005  & 3591                                 & 3580                                 & {\color[HTML]{00B050} \textbf{3509}} & 4102  & 3394 & 3235                                 & {\color[HTML]{00B050} \textbf{3172}} \\
ibm12 & 3936 & 3844 & 3940                                 & {\color[HTML]{00B050} \textbf{3699}} & {\color[HTML]{00B050} \textbf{3699}} & 3775 & 3527                                 & 3357                                 & {\color[HTML]{00B050} \textbf{3270}} & 6546  & {\color[HTML]{00B050} \textbf{6075}} & 6093                                 & 6086                                 & 6609  & 5616 & {\color[HTML]{00B050} \textbf{5433}} & 5585                                 \\
ibm13 & 1818 & 1904 & 1754                                 & 1751                                 & {\color[HTML]{00B050} \textbf{1715}} & 1817 & 1565                                 & 1563                                 & {\color[HTML]{00B050} \textbf{1562}} & 2873  & 2660                                 & {\color[HTML]{00B050} \textbf{2653}} & {\color[HTML]{00B050} \textbf{2653}} & 2927  & 2631 & 2610                                 & {\color[HTML]{00B050} \textbf{2491}} \\
ibm14 & 3346 & 3475 & 3240                                 & 3189                                 & {\color[HTML]{00B050} \textbf{3116}} & 3187 & 2878                                 & 2871                                 & {\color[HTML]{00B050} \textbf{2868}} & 5612  & 4916                                 & 4923                                 & {\color[HTML]{00B050} \textbf{4902}} & 5506  & 4428 & 4456                                 & {\color[HTML]{00B050} \textbf{4413}} \\
ibm15 & 4859 & 4720 & 4763                                 & 4769                                 & {\color[HTML]{00B050} \textbf{4582}} & 4916 & 4411                                 & 4335                                 & {\color[HTML]{00B050} \textbf{4326}} & 7379  & 6947                                 & {\color[HTML]{00B050} \textbf{6547}} & 6677                                 & 7237  & 6048 & 6010                                 & {\color[HTML]{00B050} \textbf{5963}} \\
ibm16 & 4068 & 4060 & 3758                                 & {\color[HTML]{00B050} \textbf{3746}} & {\color[HTML]{00B050} \textbf{3746}} & 4131 & 3593                                 & {\color[HTML]{00B050} \textbf{3592}} & {\color[HTML]{00B050} \textbf{3592}} & 7657  & 6636                                 & {\color[HTML]{00B050} \textbf{6544}} & 6556                                 & 7604  & 5531 & {\color[HTML]{00B050} \textbf{5515}} & 5550                                 \\
ibm17 & 5632 & 5583 & 5475                                 & 5440                                 & {\color[HTML]{00B050} \textbf{5145}} & 5663 & 4734                                 & 4716                                 & {\color[HTML]{00B050} \textbf{4649}} & 10177 & 9337                                 & 9294                                 & {\color[HTML]{00B050} \textbf{9271}} & 10213 & 8492 & 8561                                 & {\color[HTML]{00B050} \textbf{8399}} \\
ibm18 & 3183 & 2918 & 2925                                 & 2922                                 & {\color[HTML]{00B050} \textbf{2917}} & 3091 & 2748                                 & {\color[HTML]{00B050} \textbf{2745}} & {\color[HTML]{00B050} \textbf{2745}} & 7003  & 5959                                 & 5849                                 & {\color[HTML]{00B050} \textbf{5825}} & 6970  & 4569 & {\color[HTML]{00B050} \textbf{4499}} & 4537   \\
\hline
\textbf{\begin{tabular}[c]{@{}c@{}} Norm \\ Avg.\end{tabular}}    &  \cellcolor[HTML]{F2F2F2}1.034  & \cellcolor[HTML]{F2F2F2}1.043       & \cellcolor[HTML]{F2F2F2}1 & \cellcolor[HTML]{F2F2F2}0.995    & \cellcolor[HTML]{FFF2CC}\textbf{0.984} & \cellcolor[HTML]{F2F2F2}1.094  & \cellcolor[HTML]{F2F2F2}1& \cellcolor[HTML]{F2F2F2}0.993  & \cellcolor[HTML]{FFF2CC}\textbf{0.988}     & \cellcolor[HTML]{F2F2F2}1.089& \cellcolor[HTML]{F2F2F2}1        & \cellcolor[HTML]{F2F2F2}0.982    & \cellcolor[HTML]{FFF2CC}\textbf{0.975}      & \cellcolor[HTML]{F2F2F2}1.204& \cellcolor[HTML]{F2F2F2}1& \cellcolor[HTML]{F2F2F2}0.981 & \cellcolor[HTML]{FFF2CC}\textbf{0.975}   \\ \bottomrule
\end{NiceTabular}
}
\endgroup
\vspace{-.1in}
\end{table*}

\subsection{Comprehensive Performance Comparison}
\label{exp:comparison}

\subsubsection{Comparison on the Titan23 Benchmarks}
    \label{exp:titan23}
Table \ref{tab:Titan23} shows the comparison of our results with the baselines on the Titan23 benchmarks. In the scenarios with $k=4$ and imbalance=2\% and 5\%, IMPart achieves 4.6\% and 2.7\% improvement over the base partitioner KaHyPar, and 3.3\% and 1.5\% improvement over KaHyPar-E, respectively. In the scenarios with $k=10$ and imbalance=2\% and 5\%, our method IMPart achieves 2.9\% and 3.9\% improvement over the base partitioner KaHyPar, and 1.6\% and 2.3\% improvement over KaHyPar-E, respectively.

\subsubsection{Comparison on the ISPD98 Benchmarks}
    \label{exp:ispd98}
Table \ref{tab:ISPD98} shows the comparison of our results with the baselines on the ISPD98 benchmarks. In the scenarios with $k=4$ and imbalance=2\% and 5\%, IMPart achieves 1.6\% and 1.2\% improvement over the base partitioner KaHyPar, and 1.1\% and 0.6\% improvement over KaHyPar-E, respectively. In the scenarios with $k=10$ and imbalance=2\% and 5\%, our method IMPart achieves 2.5\% and 2.5\% improvement over the base partitioner KaHyPar, and 0.7\% and 0.7\% improvement over KaHyPar-E, respectively.

\subsubsection{Efficacy of the Jumping Mechanism}
    \label{exp:jump}
\begin{figure}[!t]
    \centering
    % --- Row 1 ---
    \begin{subfigure}[t]{0.49\linewidth}
        \centering
        \includegraphics[width=\linewidth]{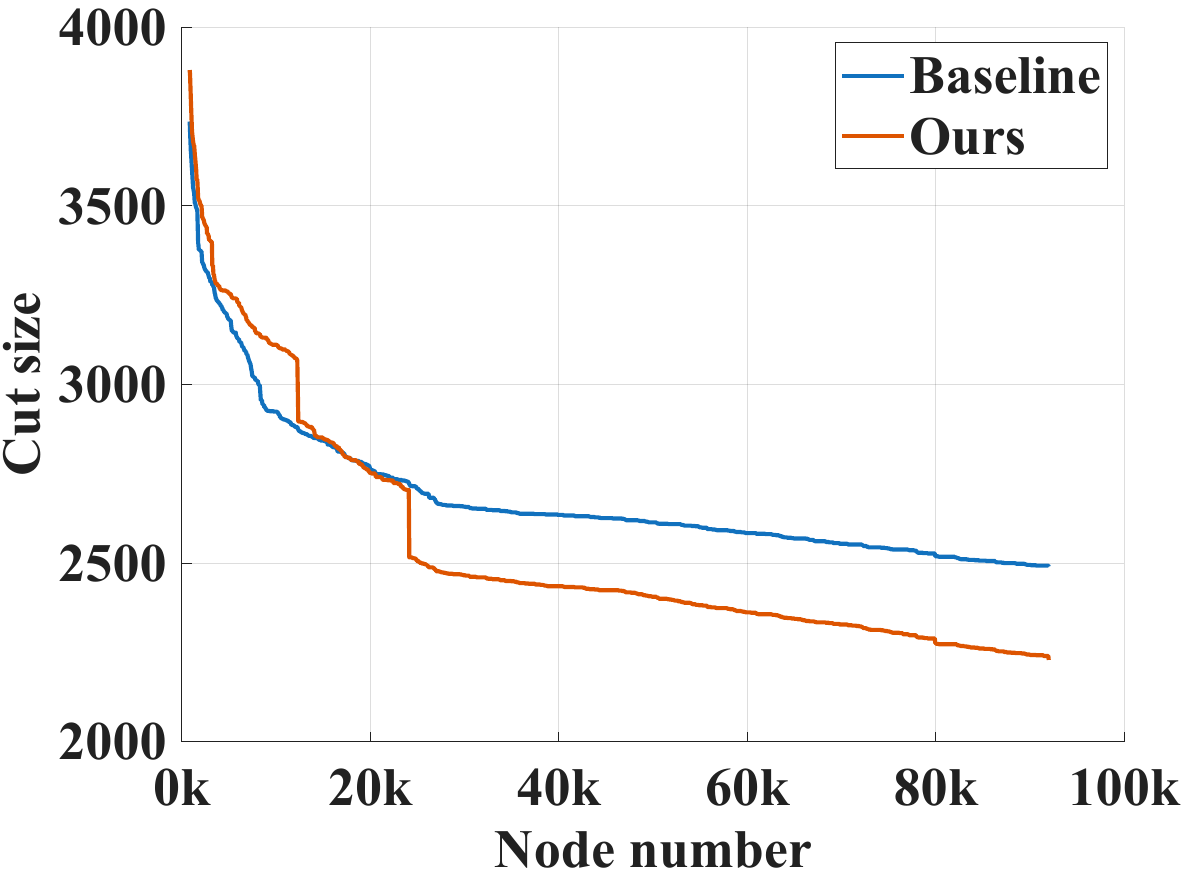}
        \caption{sparcT1\_core}
        \label{fig:subfig1}
    \end{subfigure}
    \hfill 
    \begin{subfigure}[t]{0.49\linewidth}
        \centering
        \includegraphics[width=\linewidth]{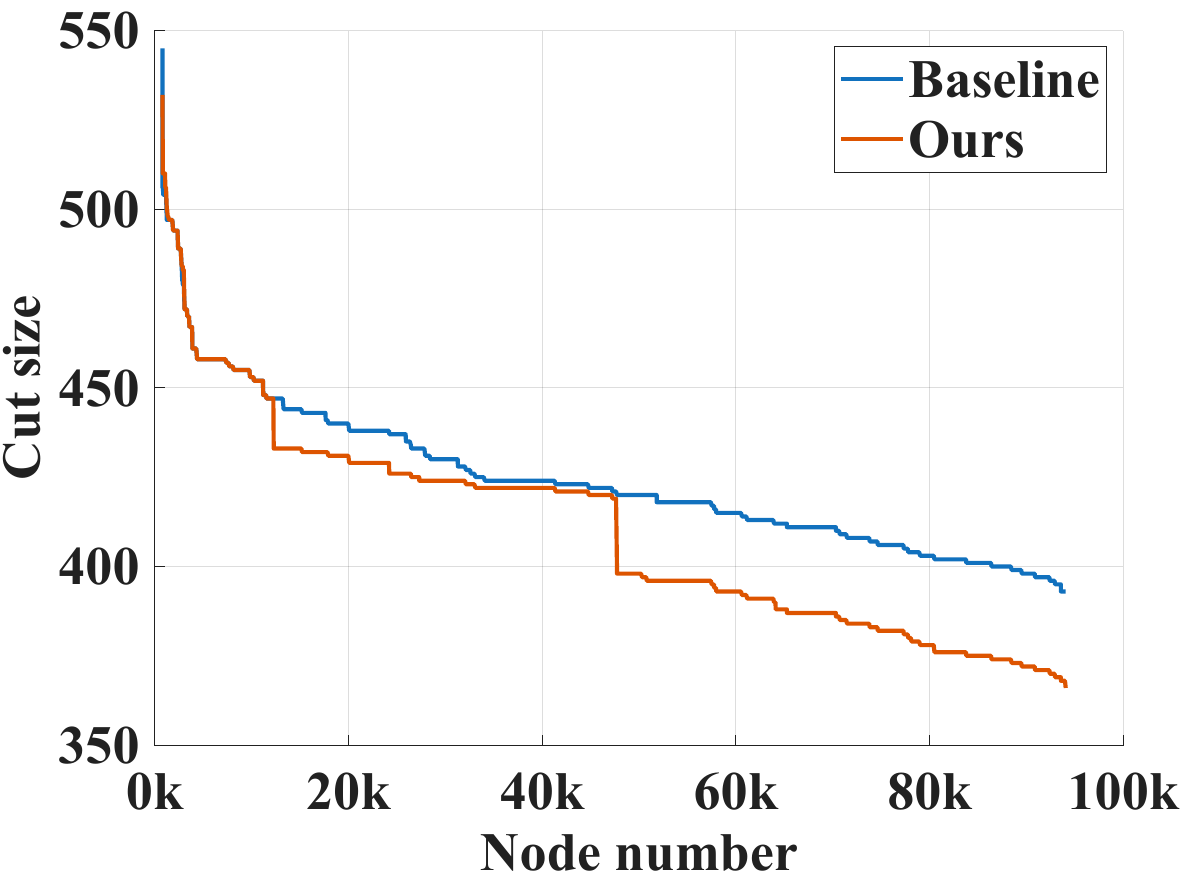}
        \caption{stereo\_vision}
        \label{fig:subfig5}
    \end{subfigure}
    \vspace{-.1in}
    \caption{The jumping mechanism in our framework: As uncoarsening proceeds and the number of nodes increases, IMPart periodically exhibits a sharp decrease in cut size.}
    \label{fig:cutsize}
    \vspace{-.1in}
\end{figure}

In this section, we demonstrate the ability to escape local optima through recombination in our framework, i.e., the \emph{jumping} mechanism. This capability is closely associated with its eventual success in achieving significantly smaller cut-sizes. Building on the KaHyPar framework, our process starts with seven distinct initial solutions (seeds -1 to 5), which then undergo continuous recombination and mutation during the uncoarsening and refinement stages. To establish a fair comparison, we configured a baseline where KaHyPar also generates seven initial solutions using the same seeds; however, each solution is processed independently through uncoarsening and refinement. For this baseline, the solution quality achieved using any of these seeds is roughly equivalent to that of a standard, single execution of KaHyPar. In Fig.~\ref{fig:cutsize}, we present a comparison on two datasets from the Titan23 benchmark suite (\texttt{sparcT1\_core} and \texttt{stereo\_vision}). It visualizes the optimization process in uncoarsening, providing clear evidence of our framework's ability to escape local optima. The optimization trajectory of our method is characterized by distinct, sharp drops in cut size, in stark contrast to the baseline, which primarily shows gradual, steady refinement. These abrupt improvements demonstrate that our recombination operator successfully identifies globally superior solutions, effectively escaping local optima that trap conventional local search methods.

\subsubsection{Scalability to Large-$k$ Partitioning}
\label{exp:scalability}
\begin{figure}[!t]
    \centering
    \includegraphics[width=0.65\linewidth]{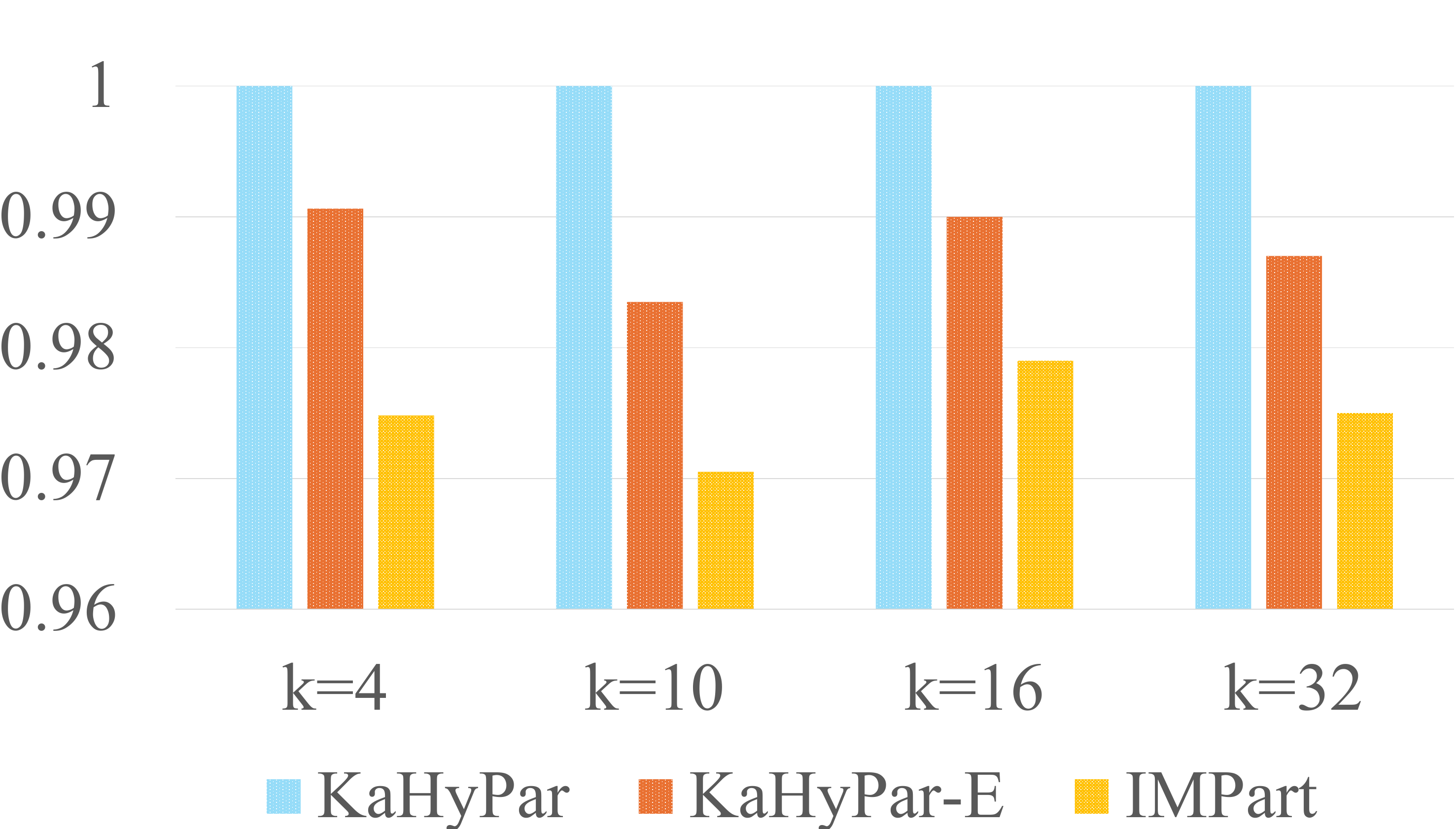}
    \vspace{-.1in}
    \caption{Normalized cut size comparison.}
    \label{fig:normalized_cutsize}
    \vspace{-.25in}
\end{figure}

In this section, we compare the normalized cut size of solutions produced by KaHyPar, KaHyPar-E, and our proposed method, IMPart, across various $k$-way partitioning scenarios. 
While the results for $k=4$ and $k=10$ are summarized from Tables~\ref{tab:Titan23} and~\ref{tab:ISPD98}, we extend this analysis to larger scenarios, specifically $k=16$ and $k=32$.
The results presented in Figure~\ref{fig:normalized_cutsize} demonstrate that our method consistently achieves superior solutions compared to both KaHyPar and the memetic algorithm KaHyPar-E.
This highlights the potential and robustness of our novel memetics-integrated multi-level framework for large-$k$-way hypergraph partitioning.

\section{Conclusion}
    \label{sec:con}
    In this paper, we propose IMPart, a memetics-integrated multi-level framework. Unlike previous memetic algorithms that employ a complete hypergraph partitioner for each recombination and mutation, we designed novel recombination and mutation operators embedded directly into the uncoarsening phase of a single multi-level partitioning process. This design makes our approach significantly more lightweight and efficient. IMPart transforms the local searches of different granularities in the traditional multi-level framework into a sophisticated, collaborative search, thereby effectively escaping local search regions. Comprehensive experimental results demonstrate that our framework consistently yields significant improvements in solution quality over state-of-the-art methods for large-$k$-way hypergraph partitioning problems.

\section{Acknowledgement}
    This work is supported by Hong Kong Research Grants Council (RGC) CRF-YCRG C6003-24Y, National Natural Science Foundation of China (NSFC) 92364102, and RGC T46-415/25-R. It was partially conducted by ACCESS – AI Chip Center for Emerging Smart Systems, supported by the InnoHK initiative of the Innovation and Technology Commission of the Hong Kong Special Administrative Region Government.

%\bibliographystyle{IEEEtran}
    %\bibliography{ref}

\bibliographystyle{ACM-Reference-Format}
    \bibliography{ref}

\end{document}